\documentclass[journal=jctcce, manuscript=article]{achemso}

\usepackage{amsmath}
\usepackage{mathtools}
\usepackage{braket}
\usepackage{multirow}
\usepackage{graphicx}
\usepackage{times}
\usepackage[table]{xcolor}
\usepackage{hyperref}
\newcommand*{\citen}{}
\DeclareRobustCommand*{\citen}[1]{%
  \begingroup
    \romannumeral-`\x 
    \setcitestyle{numbers}%
    \cite{#1}%
  \endgroup
}

\author{Qiujiang Liang}
\affiliation[]
{Department of Chemistry, The University of Hong Kong, 
Hong Kong, P.R. China}
\alsoaffiliation[]
{Hong Kong Quantum AI Lab Limited, Hong Kong, P.R. China}
\email{qliang@connect.hku.hk}
\author{Jun Yang}
\affiliation[]
{Department of Chemistry, The University of Hong Kong, 
Hong Kong, P.R. China}
\alsoaffiliation[]
{Hong Kong Quantum AI Lab Limited, Hong Kong, P.R. China}
\email{juny@hku.hk}

\title[]{Polarizable Water Model with Ab Initio Neural Network Dynamic Charges and Spontaneous Charge Transfer}

\graphicspath{{.}}

\begin{document}

\begin{tocentry}
\centering
\includegraphics[width=1.0\textwidth]{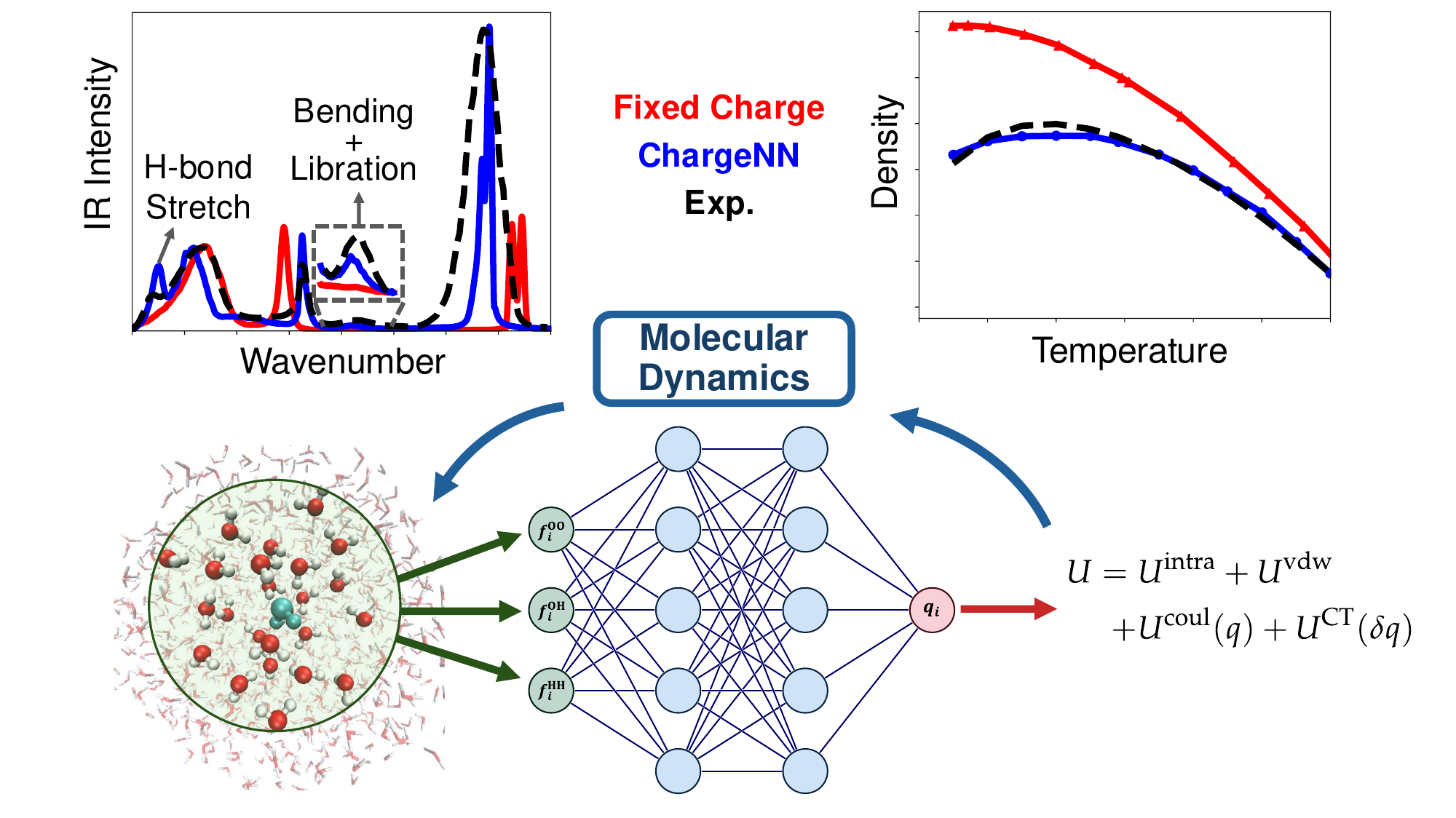}
\end{tocentry}

\begin{abstract}
Simulating water accurately has been a challenge due to the complexity of describing polarization and intermolecular charge transfer. Quantum mechanical (QM) electronic structures provide an accurate description of polarization in response to local environments, which is nevertheless too expensive for large water systems. In this study, we have developed a polarizable water model integrating Charge Model 5 atomic charges at the level of second-order M{\o}ller-Plesset perturbation theory, predicted by an accurate and transferable Charge Neural Network (ChargeNN) model. The spontaneous intermolecular charge transfer has been explicitly accounted for, enabling a precise treatment of hydrogen bonds and out-of-plane polarization. Our ChargeNN water model successfully reproduces various properties of water in gas, liquid and solid phases. For example, ChargeNN correctly captures the hydrogen-bond stretching peak and bending-libration combination band, which are absent in the spectra using fixed charges, highlighting the significance of accurate polarization and charge transfer. Finally, the molecular dynamical simulations using ChargeNN for liquid water and a large water droplet with a $\sim$4.5 nm radius reveal that the strong interfacial electric fields are concurrently induced by the partial collapse of the hydrogen-bond network and surface-to-interior charge transfer. Our study paves the way for QM-polarizable force fields, aiming for large-scale molecular simulations with high accuracy.
\end{abstract}

\section{Introduction}\label{sec1}

As an active reagent, water plays a crucial role in regulating biomolecular functions, for instance, to stabilize their structures and facilitate dynamic interactions in proteins and nucleic acids.\cite{ball2008water, levy2006water} Water also exhibits anomalous characteristics for nanomaterials, such as unique structural arrangements and fast diffusion rates within confined nanotubes compared to the bulk phase\cite{alexiadis2008molecular}. The polarization effect, which is the electron rearrangement in response to different local environments, has been shown to profoundly influence both static and dynamic properties of water.\cite{dang1998importance, piquemal2007key, koocher2015polarization, schienbein2020assessing} For example, the induced polarization by electric field is considered to be the major source of the nonadditive cooperativity in water hydrogen bonding interactions.\cite{grabowski2011covalency, mahadevi2016cooperativity} Moreover, the considerable electron transfer between water molecules enhances the proton donor-acceptor orbital interactions and hence the covalency,  directionality as well as relevant spectroscopic properties of water hydrogen bonds.\cite{elgabarty2015covalency} In recent years, microdroplets have been shown to accelerate chemical reactions by up to 10 million times compared to bulk solutions, widely attributed to the strong electric fields formed on their surfaces. One promising origin of the interfacial electric fields is the large number of H$_{2}$O$^{\delta+}$/H$_{2}$O$^{\delta-}$ radical pairs on the microdroplet surface created by the intermolecular charge transfer\cite{vacha2012charge, qiu2021reaction, qiu2022spontaneouswater, ben2022electric}. A recent second-order M{\o}ller-Plesset perturbation theory (MP2) based molecular dynamics (MD) study indicates that substantial and inhomogeneous interfacial water charge transfer dramatically promotes the reactions of Criegee intermediates with water. \cite{liang2023water}

 Atomic partial charge is a qualitative representation of the environment-dependent electron distribution for understanding polarization effects and rationalizing chemical phenomena.\cite{hedegaard2009partial, cheng2022stability,de2002atomic,nikolova2019atomic,kiryanov2016prediction}. Nevertheless, partial charge is neither experimentally nor theoretically observable. In quantum chemistry, atomic charges vary significantly with different electron density partition methods, such as Mulliken\cite{mulliken1955electronic}, L{\"o}wdin\cite{lowdin1950non}, natural population analysis (NPA)\cite{foster1980natural}, quantum theory of atoms in molecules (QTAIM)\cite{bader2004atomic}, Hirshfeld\cite{hirshfeld1977bonded}, etc. Notably, the Hirshfeld population analysis precisely describes the density redistribution during bonding.\cite{hirshfeld1977bonded} Furthermore, the charge model 5 (CM5) corrects the underestimated Hirshfeld charges to achieve accurate dipole moments\cite{marenich2012charge}, which has been shown to provide accurate electronic distribution and to be insensitive to the choice of basis sets and theories.\cite{wiberg2018atomic} Calculating accurate electron densities to produce partial charges is computationally demanding using density functional theory (DFT) or higher level correlated wavefunction methods (e.g. MP2 and coupled cluster singles and doubles (CCSD)), making it cost-prohibitive for large molecular systems. In the past decades, efforts have been made to develop machine learning (ML) charges by mapping the atomistic local environment descriptors onto QM reference atomic charges.\cite{houlding2007polarizable, handley2009optimal, bleiziffer2018machine, martin2019contradrg, wang2020fast, wang2020atomic, kancharlapalli2021fast,han2023incorporating, korolev2020transferable, nebgen2018transferable, sifain2018discovering, unke2019physnet, kato2020high, gallegos2022nnaimq, gallegos2023developing, gallegos2024explainable, raza2020message, wang2021deepchargepredictor, wang2021deepatomiccharge, metcalf2020electron, jiang2022out, lehner2023dash, wang2024espalomacharge} However, these charge models were mainly tested on single molecules and small clusters, casting doubt on their accuracy when applied to larger molecular systems. A recent study reports a kernel ridge regression (KRR) model accurately predicts iterative Hirshfeld atomic charges\cite{bultinck2007critical} of water molecules in bulk liquid represented by QM/molecular mechanical (QM/MM) water clusters.\cite{han2023incorporating} Remarkably, the hydrogen-bond stretch infrared (IR) peak is successfully retrieved using dynamic ML charges, whereas this weak peak disappears for fixed charges, marking the importance of polarization and intermolecular charge transfer in simulating liquid water. \cite{han2023incorporating} However, this model has several limitations. Firstly, the training and testing were conducted with an identical system (23 QM water molecules with 1977 MM water molecules), leaving the model’s generalizability to larger QM regions uncertain. Secondly, the charge model was specifically tailored for water in bulk solution, which makes it inadequate for predicting charges for interfacial water molecules that experience more complex local environments. Finally, while ML charges were used to compute dipole moments and IR spectra using trajectories prepared with TIP4P/2005, they have not been integrated into a water model for simulating polarization in molecular dynamics.

Simulating water has been challenging for decades.\cite{demerdash2018advanced} Since the introduction of the rigid non-polarizable water models with fixed charges\cite{berendsen1981interaction, jorgensen1983comparison, 
mahoney2000five}, they have been substantially improved by
refining parameters \cite{horn2004development, 
abascal2005potential, abascal2005general, izadi2014building, 
izadi2016accuracy}, introducing intramolecular flexibility\cite{wu2006flexible, 
wang2014building, habershon2009competing}, and integrating
implicit polarizability\cite{berendsen1987missing}. Moreover, 
tremendous efforts have been made to recover the explicit
polarization effects. Some polarizable models incorporate induced multipoles in response to the local electric fields using atomic polarizability tensors \cite{caldwell1995structure, shi2013polarizable, laury2015revised, liu2019amoeba+, liu2019implementation, burnham1999parametrization, burnham2002development, fanourgakis2008development, heindel2024completely, houlding2007polarizable, handley2009optimal, symons2021dl_fflux, hughes2020fflux, symons2022application, brown2023application} or Drude
oscillators\cite{lamoureux2003simple, lamoureux2006polarizable, wang2022flexible, yu2013six, xiong2022fast}.
In contrast, the charge-flow models
enable geometry-dependent charge redistribution\cite{rappe1991charge, rick1994dynamical, oliferenko2006atomic, 
patel2004charmm, leven2021recent, 
chen2017accurate, banks1999parametrizing, 
nistor2006generalization, muser2012chemical, 
dapp2013towards, morita1998molecular, 
nakano2010wave, verstraelen2013acks2, 
verstraelen2014direct, holt2007charge, holt2009charge}. Although
charge-flow polarizable models appear enticing due to their
explicit incorporation of charge exchange between atoms, they
suffer from several deficiencies, including high computational
costs associated with iteratively updating charges, non-linear
computational scaling of the polarizability with system sizes, 
inadequate description of intermolecular charge transfer, and
a lack of out-of-plane polarization.\cite{jensen2022using, 
jensen2023unifying} Furthermore, despite extensive fine-tuning
of the charge functional parameters to fit the DFT electrostatic
potentials, the absence of spontaneous quantum mechanical
information during the solution of dynamical charges may result
in questionable electron density arrangements. Beyond the empirical force fields, energy potentials from first-principles
 for large water systems can be built by employing many-body expansion (MBE) for potentials of small water
fragments obtained through fitting with functionals or machine learning.\cite{bukowski2007predictions, bukowski2008polarizable, huang2006ab, wang2011flexible, wang2011ab, wang2009full, medders2013critical, babin2012toward, babin2013development, babin2014development, medders2014development, 
behler2007generalized, bartok2010gaussian, bartok2013representing, rupp2012fast, unke2021machine, yu2023status}
Nevertheless, constructing an accurate potential requires a tremendous amount of training data to cover the vast chemical
space. The applicability of potentials is severely limited due to their system-specific nature that does not generalize well to unseen molecules. Furthermore, important electronic properties such as electron density and multipole moments are absent in the potentials. Notably, the QM-polarizable force fields that incorporate instantaneous QM-level polarizability enabled by machine-learning, such as FFLUX using machine learned QTAIM atomic multipoles, have been successfully applied to various chemical systems.\cite{houlding2007polarizable, handley2009optimal, thacker2018towards, symons2021dl_fflux, hughes2020fflux, symons2022application, brown2023application}

In this study, we present a novel QM-polarizable water model that incorporates dynamical atomic charges and charge transfer exclusively trained on QM data acquired at the \textit{ab initio} MP2 level of theory. The charges are produced using a deep neural network charge model called ChargeNN.  A set of Interaction Classified Functions (ICFs) was employed as the ChargeNN features, explicitly capturing the types of atomic interactions to accurately depict local environments. Trained on diverse (H$_2$O)$_{25}$ structures sampled from MD trajectories, the ChargeNN is capable of accurately predicting charges for much larger water clusters, for example, (H$_2$O)$_{190}$, as compared to computed MP2 charges. The ChargeNN has also been integrated into a water model for dynamical polarization simulations. The intermolecular charge transfer has been explicitly accounted for using machine-learned pair-wise water charge transfer, which effectively captures the out-of-plane polarization. We have implemented analytical energy gradients in the periodic boundary condition (PBC) within the shared-memory message passing interface (MPI) framework for efficient MD simulations. Our results validate the capability of ChargeNN-derived water model to reproduce a variety of water properties in excellent agreement with experimental data, for instance, the model generalizes well to a few thousand water molecules in unit cell for predicting the liquid water properties. Finally, the statistical analysis of liquid water and large water microdroplets using ChargeNN reveals that the partial breakage of hydrogen-bond networks, along with water charge transfer, concurrently promotes the layer electric fields at the air/water interface. The ChargeNN water model is thus of broad applicability in studying condensed ice, liquid water, nano-phase, large microdroplets, and solvent interactions with materials and biomolecules. 

\section{Methods}
\subsection{Charge Model 5}
Charge Model 5 (CM5)\cite{marenich2012charge} based on  Hirshfeld population analysis (HPA)\cite{hirshfeld1977bonded} was implemented for third-order many-body-expansion orbital-specific-virtual MP2 (MBE(3)-OSV-MP2) to investigate the water polarization and charge transfer in response to the variation in the local environment. The Hirshfeld atomic charges ($q^{\mathrm{HPA}}_{i}$) are obtained by integrating the electron density within the region surrounding each atom,
 \begin{equation}
     q^{\mathrm{HPA}}_{i} = Z_{i} - \int \mathrm{d}\mathbf{r}\frac{\rho^{0}_{i}(\mathbf{r})}{\sum_{j}\rho^{0}_{j}(\mathbf{r})}\rho(\mathbf{r}),
 \end{equation}
 where $Z_{i}$ is the nuclear charge. $\rho^{0}(\mathbf{r})$ and $\rho(\mathbf{r})$ denote the electron densities of the promolecule and real molecule at position $\mathbf{r}$, respectively. In contrast to the hard-boundary partitioning approaches like QTAIM\cite{bader1990quantum, bader1991quantum}, where electron density at each grid point is attributed to a single atom, the boundary-less Hirshfeld population analysis allows electron density to contribute to multiple atoms.  It provides a clear partitioning of the electron density and accurately describes the density redistribution from promolecule to real molecule during bonding. Nevertheless, Hirshfeld charges are significantly underestimated due to the Hirshfeld weighting factor, which causes the atomic population to closely resemble that of the isolated atom. \cite{bultinck2007critical}
 
\begin{figure}[H]
  \includegraphics[width=\linewidth]{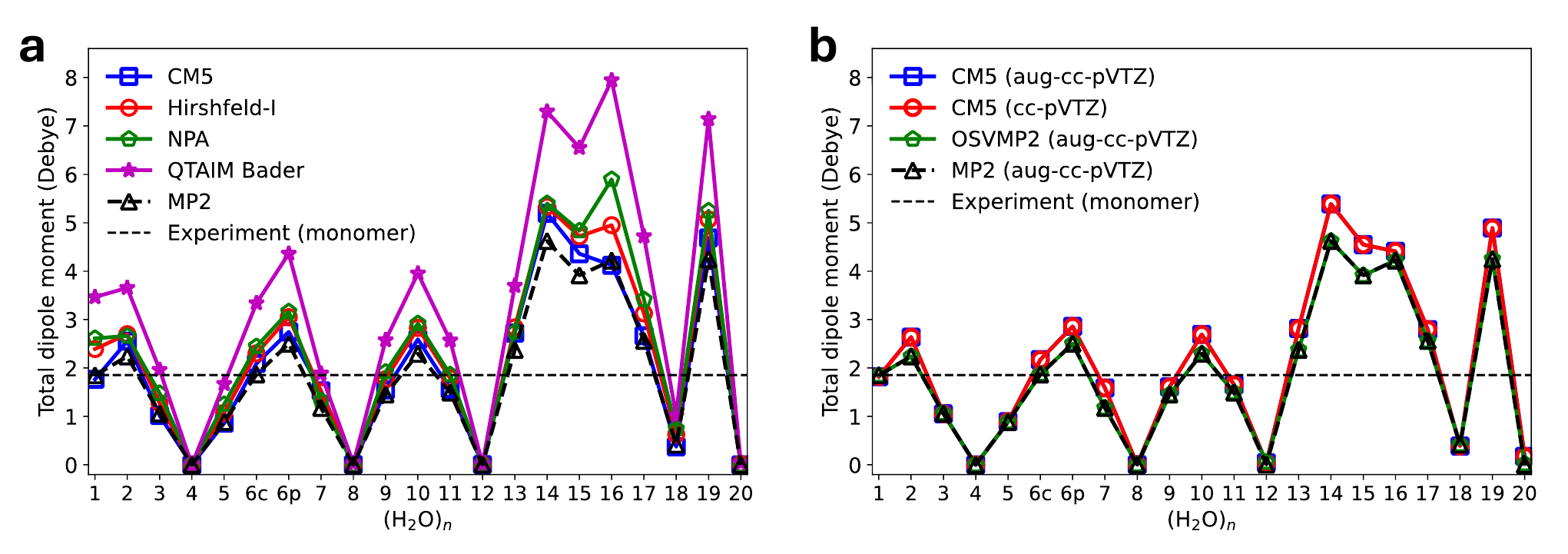}
  \centering
  \caption{\textbf{Water dipole moments derived from charges} (\textbf{a}) Water dipole moments obtained from CM5, Hirshfeld-I\cite{bultinck2007critical}, NPA\cite{reed1985natural} and QTAIM charges, using electron density of MP2/aug-cc-pVTZ computed with Gaussian 16\cite{g16}. CM5 and NPA charges were obtained with Gaussian 16, while Multiwfn\cite{lu2012multiwfn, lu2024comprehensive} was used to compute Hirshfeld-I and QTAIM charges. (\textbf{b}) Water dipole moments obtained from CM5 charges, computed with MBE(3)-OSV-MP2 using aug-cc-pVTZ and cc-pVTZ basis sets. Experimental structure of water monomer and structures from Cambridge water clusters\cite{maheshwary2001structure} were used for this test. (H$_2$O)$_{6\mathrm{c}}$ and (H$_2$O)$_{6\mathrm{p}}$ refer to cage and prism conformers of water hexamers, respectively. Experimental dipole moment of the water monomer is from ref.\citen{clough1973dipole}.}
\label{fig:dip_compare}
\end{figure}

 By introducing a unified set of parameters optimized by fitting to the experimental or DFT dipole moments of 614 molecules, the CM5 model maps the underestimated Hirshfeld charges onto a new set of charges, providing a more accurate description of the electrostatic potential:
 \begin{equation}
 \begin{split}
     q^{\mathrm{CM5}}_{i} ={}& q^{\mathrm{HPA}}_{i} + \sum_{j\neq i}A_{ij}B_{ij}, \\
     B_{ij} ={}& \text{exp}\left[-\alpha (r_{ij} - R_{Z_{i}} - R_{Z_{j}})\right],
 \end{split}
 \end{equation}
 where $A_{ij}$ and $\alpha$ are the CM5 model coefficients to be optimized. Constant $R_{Z}$ denotes the atomic covalent radius. CM5 has been shown to
deliver precise electronic charge distributions and be insensitive to different theoretical levels and basis sets.\cite{marenich2012charge, wiberg2018atomic}. As demonstrated in Figure \ref{fig:dip_compare}\textbf{a}, CM5 charges outperform iterative Hirshfeld (Hirshfeld-I), NPA, and QTAIM charges in reproducing the dipole moments of water clusters containing 1-20 molecules. Furthermore, CM5 exhibits  numerically consistent dipole moments between cc-pVTZ and aug-cc-pVTZ basis sets, as shown in Figure \ref{fig:dip_compare}\textbf{b}.
The CM5 charges in this work were obtained from the MBE(3)-OSV-MP2/cc-pVTZ relaxed density matrix. 

\subsection{Deep Neural Networks for Atomic Charges}
Tremendous efforts have been made to design descriptors of local atomic environments for building machine learning models of energy potentials and atomic properties.\cite{behler2007generalized, behler2011atom, behler2016perspective, bartok2010gaussian, bartok2013representing, rupp2012fast} For instance, the well known radial ($G_{i}^{\text{rad}}$) and angular ($G_{i}^{\text{ang}}$) atom-centered symmetry functions (ACSFs)\cite{behler2011atom} write 
\begin{equation}
\begin{split}
    G_{i}^{\text{rad}} = {}&\sum_{j} e^{-\eta (r_{ij} - r_{s})^{2}} s_{c}(r_{ij}), \\
    G_{i}^{\text{ang}} ={}& 2^{1-\zeta} \sum_{\substack{j,k \neq i }} \left( 1 + \lambda \cos \theta_{ijk} \right)^{\zeta} \cdot e^{-\eta \left( r_{ij}^{2} + r_{ik}^{2} + r_{jk}^{2} \right)} \\ 
    {}& \cdot s_{c} \left( r_{ij} \right) s_{c} \left( r_{ik} \right) s_{c} \left( r_{jk} \right), 
\end{split}
\end{equation}
where $\eta$, $r_{s}$, $\zeta$ and $\lambda$ are tunable parameters. $s_{c}$ denotes the cutoff function that enforces a smooth transition of atoms $j$ entering and exiting the selection region for the atom $i$

\begin{equation}
    s_{c}(r_{ij}) =
  \begin{cases}
  0.5\left[ \text{cos}\left( \frac{\pi r_{ij} }{r_{c}} \right) + 1 \right], & r_{ij}\le r_{c}\\
  0,       & r_{ij} > r_{c}
\end{cases}.
\label{eq:sc}
\end{equation}
However, ACSF descriptors omit the dependence of the atomic interactions on element type, which can significantly impact training performance and data efficiency.

To incorporate the information of atomic interaction types, we propose an interaction classified function (ICF) to describe the interaction-specific ($\alpha\beta$) contributions to the charge of atom $i$ of the atom type $\alpha$, which satisfies translational, rotational and permutational invariance but requires no angular ACSF parameterization for angularly independent atomic charges:
\begin{equation}
\begin{split}
  f^{\alpha\beta}_{i} = {}&\sum_{j\in\beta} e^{- \frac{r_{ij}^{2}}{k_{f}}} s_{c}(r_{ij}),\textrm{ where } i\in \alpha.\\
\end{split}
\label{eq:icf}
\end{equation}
In the above formula, a single parameter $k_{f}$ controls the interaction decay with respect to the atomic distance, effectively alleviating the heavy process of the parameter tuning. Despite their simplicity, the radial ICFs have been found to sufficiently describe local interactions needed for reproducing accurate atomic charges with excellent data efficiency (see section 3.1).
To determine the atom selection cutoff ($r_{c}$) in eq. \ref{eq:sc}, we tested the charge convergence for both bulk water and air/water interface, and found that 24 surrounding water molecules sufficiently form the local environment to produce charges comparable to those surrounded by 127 molecules, as demonstrated in Figure S1. The selected cutoff $r_{c}$ was accordingly set to be 4.4 {\AA}, which is the minimum radius of multiple (H$_{2}$O)$_{25}$ droplets extracted from well-equilibrated liquid water boxes. 

The ICF feature set for the charge of atom $i$ in the molecule $I$ reads
\begin{equation}
  \mathbf{F}_{i_{I}} = \{f_{i_{I}}^{\alpha_{I}\beta_{I}}, f_{i_{I}}^{\alpha_{I}\beta_{J}}, f_{j_{I}}^{\alpha_{I}\beta_{J}}\} \label{eq:fea_atom}.
\end{equation}
where the intra- ($f_{i_{I}}^{\alpha_{I}\beta_{I}}$) and intermolecular ICFs ($f_{i_{I}}^{\alpha_{I}\beta_{J}}$) are separately placed for detailing interaction types. In addition, the intermolecular ICFs of other atoms in the same molecule ($f_{j_{I}}^{\alpha_{I}\beta_{J}}$) are also appended to the feature set to account for the perturbation arising from the interactions for the neighboring atoms, which notably improves the prediction accuracy. Three decay parameters $k_{f}$ were employed for each ICF in eq. \ref{eq:fea_atom} to simulate varying interactions with the atomic distance, as illustrated in Figure S2. To better handle the distinct charge ranges and distributions, two separate neural networks were specifically trained for the charges of oxygen and hydrogen.

Although the electrostatic interaction, modeled by Coulomb potential with flexible atomic charges, implicitly accounts for environment-dependent charge flows, it is substantially underestimated in the absence of the intermolecular charge transfer (CT) contribution. For recovering the missing interactions, an extra neural network is used to predict intermolecular CT between water pairs ($IJ$) by supplying a feature set composed of intermolecular ICFs:
\begin{equation}
\begin{split}
    \mathbf{F}_{IJ} ={}& \{f^{\alpha_{I}\beta_{J}}\},\\
    f^{\alpha_{I}\beta_{J}} = {}&\sum_{\substack{i\in \alpha_{I} \\ j\in\beta_{J}}} e^{- \frac{r_{ij}^{2}}{k_{f}}} s^{\mathrm{CT}}_{c}(r_{ij}).\\
\end{split}
\end{equation}
We examined the decay of the CT with the atomic distance, as the decay of the CT energy in eq. \ref{eq:water_model} specifically depends on CT. Figure S3 shows that CT deteriorates to only $\sim$$10^{-5}$ e at the hydrogen bond distance of 5.5 {\AA} for a water dimer. Considering that CT would also be hindered by the water molecules filled between distant water pairs, 
5.5 {\AA} is sufficiently large for the CT cutoff.  

The CM5 atomic charges of (H$_2$O)$_{25}$ were utilized to train the neural networks. Reference charges were computed using the low-scaling MBE(3)-OSV-MP2 method\cite{liang2021third} with the cc-pVTZ basis set, as MP2 has been shown to accurately predict the energetic and electronic properties of water\cite{miliordos2013optimal, miliordos2015accurate, han2021determining}. To benchmark the charge model, charges were prepared for 20000 (H$_2$O)$_{25}$, 200 (H$_2$O)$_{32}$, 50 (H$_2$O)$_{64}$, 10 (H$_2$O)$_{64}$ and 1 (H$_2$O)$_{190}$ cluster conformations that were randomly sampled from the equilibrated MD trajectories using SWM4-NDP with harmonic bonds and angles on OpenMM\cite{eastman2013openmm, huang2018molecular}. The $NVT$ trajectories were simulated in the gas phase at 298.15 K, with a time step of 1 femtosecond.

Four hidden layers are employed to effectively capture intricate relationships between features and atomic charges, which also mitigates overfitting issues. The Swish function\cite{ramachandran2017searching} is adopted as the activation function:
\begin{equation}
  A(x) = \frac{x}{1 + e^{-\beta x}}.
\end{equation}
The Swish function is differentiable and ensures smooth charge functions for new molecular configuration, which avoids discontinuity on the potential energy surface. Following multiple trials, we discovered that utilizing $\beta=3$ allows $A(x)$ to retain good linearity around $x=0$, enhancing the learning performance while circumventing erratic charge functions caused by overfitting. The mean squared error (MSE) was chosen as the loss function. Among the 20,000 configurations of (H$_2$O)$_{25}$, 8,000 were randomly selected for training, with 50 used as the validation data set. The remaining data points were used for evaluating the prediction performance. After extensive testing, we applied the ``early stopping'' technique with 200 epochs to ensure satisfactory loss convergence and to prevent overfitting.

\subsection{Water model with ChargeNN charges}
The potential energy functional of the 3-site flexible water model incorporating ChargeNN charges and charge transfer contributions is expressed as a sum of the intramolecular energy ($U^{\mathrm{intra}}$) and the intermolecular energy ($U^{\mathrm{inter}}$),
\begin{equation}
\begin{split}
  U^{\mathrm{intra}} {}&= \sum_{I}\left[\sum_{i}\frac{k_{b}}{2}(r_{\text{O}_{I}\text{H}_{i_{I}}} - r^{0}_{\text{OH}})^{2} + 
  \frac{k_{a}}{2} (\theta_{I} - \theta^{0})^{2}\right], \\
U^{\mathrm{inter}} {}&=
   \sum_{i<j}\varepsilon_{ij} \left[7\left(\frac{\sigma_{ij}}{r_{ij}} \right)^{9} - 5\left(\frac{\sigma_{ij}}{r_{ij}} \right)^{6} \right] \\
  {}& + \sum_{i<j} k_{e}\frac{q_{i}q_{j}}{r_{ij}} - \sum_{I<J}k_{c} \frac{\delta q_{IJ}^{2}}{\delta q_{IJ}^{2} + 3(\delta q^{0})^2}.
\end{split}
\label{eq:water_model}
\end{equation}
In the formula for intramolecular energy, $k_b$ and $k_a$ represent the coefficients of bond stretching and bending, respectively. $r_{\text{O}_{I}\text{H}_{i_{I}}}$ denotes the O-H bond length, while $\theta_{I}$ refers to the H-O-H angle of the molecule $I$, with $r^{0}_{\text{OH}}$ and $\theta^{0}$ being the equilibrium values. For the intermolecular energy, $\epsilon_{ij}$ and $\sigma_{ij}$ are standard 12-6 Lennard-Jones (LJ) parameters. $k_{e}$ and $k_{c}$ pertain to coulomb constant and charge-transfer constant, respectively. $q_{i}$ denotes the charge of atom $i$. $\delta q_{IJ}$ represents the charge transferred between molecules $I$ and $J$, with the parameter $\delta q^{0}$ as the reference CT that yields the strongest CT force. 

We observed that atomic charges exhibit significant sensitivity to stretching bonds and bending angles, necessitating the addition of flexibility in the intramolecular forces. Moreover, this flexibility has been reported to improve predicted water properties, and also permits recovering intramolecular vibrational spectra.\cite{wu2006flexible, wang2022flexible} Consequently, we adopted harmonic bonds and angles for better description of the dynamical behavior of the water molecules. The equilibrium bond length $r^{0}_{\text{OH}}$ and angle $\theta^{0}$ were tuned to ensure that the average values in the liquid phase align with experimental measurements at 298.15 K and 1 atm. The coefficients of bond length ($k_{b}$) and angle ($k_{a}$) were adjusted to produce accurate positions of the bond stretching and bending peaks in the IR spectrum.

The electrostatic energy is calculated using ChargeNN charges. To facilitate simulations under periodic boundary conditions (PBC), Ewald summation is employed, utilizing the minimum image convention to construct the local environment for charge generation. It is important to note that the total charge of a cluster or unit cell could be non-zero due to the accumulative error of atomic charges. To ensure a zero net charge of the whole system, all atomic charges must be subtracted by a small constant ($\Delta q = \sum_{i}q_{i}/N$, where $N$ is the number of atoms) of approximately $\pm$0.0001 e. Additionally, non-homogeneous correction can be accomplished by considering the weights of atomic charges\cite{kancharlapalli2021fast} or integrating an explicit penalizing term to enforce electroneutrality into the loss function. There have been attempts to include charge transfer interactions with empirical charge transfer.\cite{liu2011insights, soniat2012effects,rick2023effects,kumar2017charge} For our model that uses CT obtained quantum mechanically, the charge transfer term formulates a smooth ratio function of squared CT. The formula offers several merits, for instance, it is symmetric and the CT energy converges to zero in the absence of CT, preventing abrupt changes in potential energy around the CT cutoff. Additionally, the formula effectively captures the rapid intensification of CT interactions when initial charges are small, as well as the gradual attenuation due to the electron migration from equilibrium. Moreover, since the inter-water CT occurs away from the nuclear sites, the CT term naturally accounts for out-of-plane polarization, resulting in more accurate water conformations. 

As the inclusion of the CT term strengthens the short-range interactions, the traditional 12-6 LJ potential for Van der Waals (VDW) forces yielding a steep repulsive wall constrains inter-water distances to a compact range, which leads to an excessively high first peak in the oxygen-oxygen radial distribution function (RDF). Instead, we employ a 9-6 functional form with minor revisions to the original LJ formula, retaining the 12-6 LJ parameters $\varepsilon$ and $\sigma$ for compatibility with the existing force fields. In the process of the parameter tuning, the VDW equation was written as $U_{ij}^{\mathrm{vdw}} = A_{ij}/r^{9}_{ij} - B_{ij}/r^{6}_{ij}$ for convenience. The intermolecular parameters $k_{c}$, $k_{d}$, $A_{ij}$ and $B_{ij}$ were optimized for the accurate heat of vaporization, density and radial distribution functions (RDF) for liquid water at the ambient conditions. All parameters can be found in Table S1.

\subsection{Implementation}
\begin{figure}[H]\includegraphics[width=0.8\linewidth]{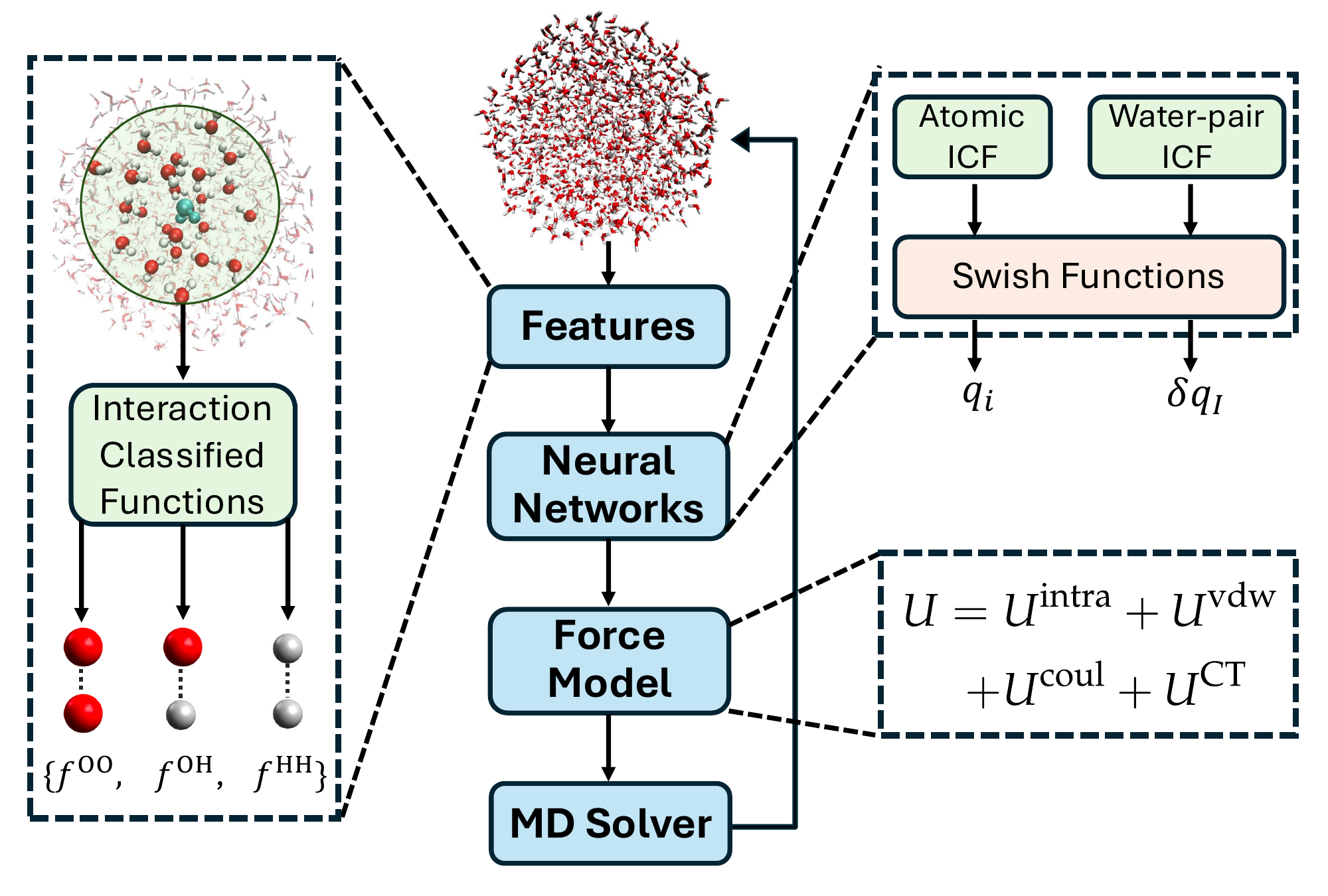}
  \centering
  \caption{\textbf{Implementation flowchart.} Schematic illustration for the implementation of the polarizable water model using the charges produced by the ChargeNN model. }
  \label{fig:flow}
  \end{figure}
Figure \ref{fig:flow} illustrates the implementation of the ChargeNN water model for molecular dynamics (MD) simulations. In each MD step, atomic coordinates adjusted by the MD solver are utilized to compute the ML features interaction-classified functions $\mathbf{F}$ and their analytical derivatives with respect to nuclear positions $\mathrm{d}\mathbf{F}/\mathrm{d}\mathbf{R}$. The ML features are then fed into the pre-trained neural networks for the predictions of atomic charges and charge transfer. By employing auto-differentiation through back-propagation, the gradients of the charges concerning the features $\mathrm{d}q/\mathrm{d}\mathbf{F}$ can be conveniently evaluated. Subsequently, the predicted atomic charges and charge transfer are used to determine the potential energy as described in eq. \ref{eq:water_model}, where total gradients are obtained using the chain rule:
\begin{equation}
    \frac{\mathrm{d}U}{\mathrm{d}\mathbf{R}} = \frac{\partial\bar{U}}{\partial\mathbf{R}} + \frac{\partial U}{\partial q} \cdot \frac{\mathrm{d}q}{\mathrm{d}\mathbf{F}} \cdot\frac{\mathrm{d}\mathbf{F}}{\mathrm{d}\mathbf{R}}.
\end{equation}
where $\partial\bar{U} / \partial\mathbf{R}$ denotes the derivative of all terms that are explicitly dependent on the nuclear coordinates $\mathbf{R}$. Finally, the energy and gradients are passed to the MD solver to update the positions. 

The ChargeNN program was written in Python with double precision, integrated with Tensorflow for handling neural networks. The algorithm has been parallelized based on Message Passing Interface (MPI) standard of version 3. The shared memory windows built in MPI-3's remote memory access module enable significantly lower-latency data communications compared to traditional point-to-point communications.

\section{Results and discussion}\label{sec5}
\subsection{Performance of the charge model}
  \begin{figure}[H]
  \includegraphics[width=\linewidth]{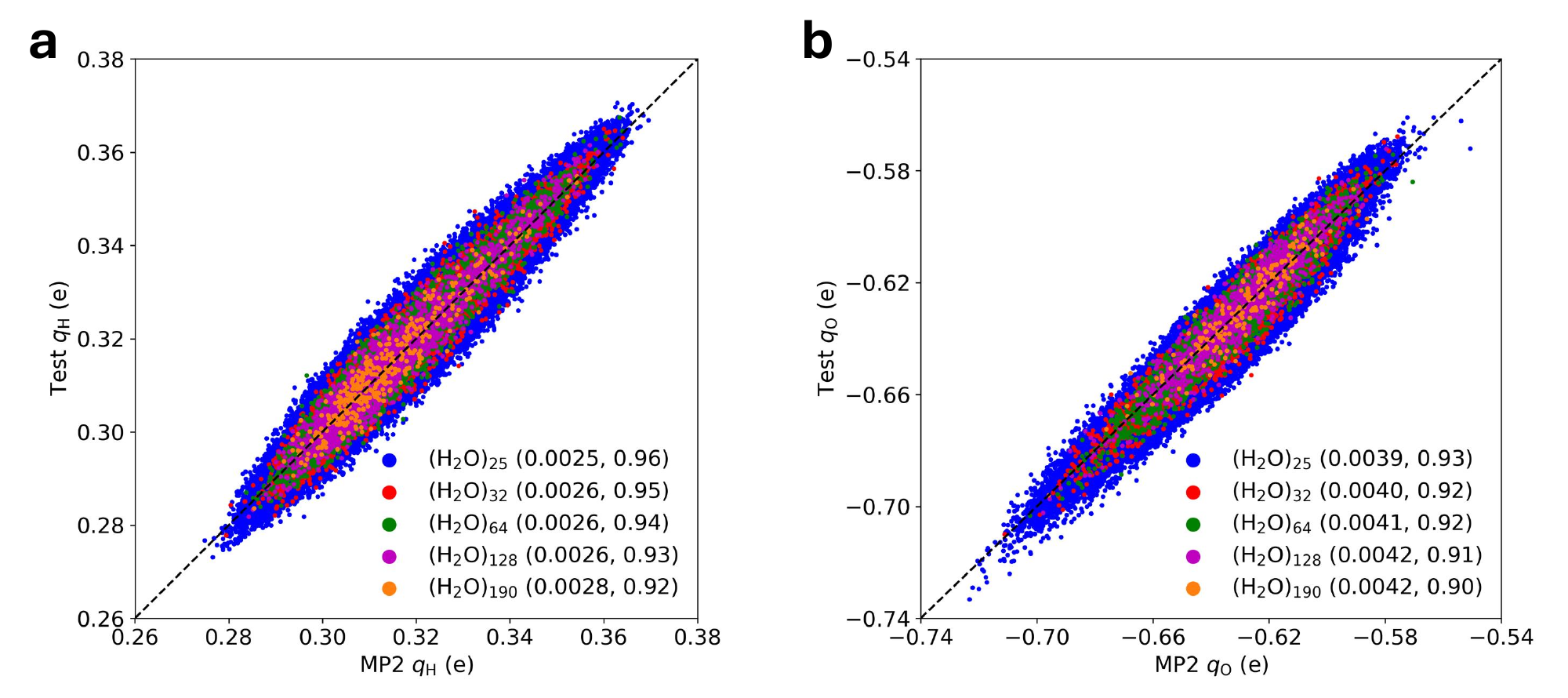}
  \centering
  \caption{\textbf{Benchmark results for the ChargeNN model.} Comparisons of ChargeNN charges and MP2 charges for \textbf{a}. hydrogen and \textbf{b}. oxygen atoms of testing water clusters containing 25, 32, 64, 128 and 190 molecules. The charge models were trained with (H$_2$O)$_{25}$. The MAEs in elementary charge (e) and coefficients of determination ($R^2$) are given in the right brackets.}
  \label{fig:ml_chg}
  \end{figure}

The ChargeNN model demonstrates exceptional data efficiency, requiring only 8000 training (H$_2$O)${_{25}}$ clusters to achieve converged mean absolute errors (MAEs) for the remaining testing clusters, as shown in Figure S4. According to Figure \ref{fig:ml_chg}, when trained on 8000 (H$_2$O)$_{25}$, the charge model yields testing MAEs of 0.0025 e for hydrogen and 0.0039 e for oxygen within the same system. These values are markedly lower than the respective MAEs of 0.0036 e and 0.0045 e for (H$_2$O)$_{23}$ predicted by a recent KRR water charge model trained with the same number of data points\cite{han2023incorporating}. 
In addition, our model shows exceptionally high prediction reliability with the coefficients of determination ($R^2$) of 0.96 for hydrogen and 0.93 for oxygen, outperforming the 0.88 reported for the KRR model\cite{han2023incorporating}. Notably, Figure \ref{fig:ml_chg} shows that the ChargeNN model considerably improves the transferability, only with a slow deterioration in accuracy as the water cluster size increases. For example, the models trained with (H$_2$O)$_{25}$ make accurate prediction of the charges for much larger (H$_2$O)$_{190}$, yielding MAEs of 0.0028 e for hydrogen and 0.0042 e for oxygen, both with $R^2$ values above 0.9. Figure S5 shows that predicting molecular charges by summing the ChargeNN atomic charges within each molecule is less satisfactory, with $R^2>0.92$ and marginally larger MAEs of $\sim$0.007 e due to the accumulative errors. This error can be improved in future by imposing constraints on the loss function to penalize the predicted molecular charges. For assessing ChargeNN's performance in predicting charge transfer, we prepared charge transfer of 40000 water pairs extracted from the MD trajectory of (H$_2$O)$_{25}$. The ChargeNN trained on 20000 water pairs achieves an MAE of only 0.001 e and $R^2$ of 1.00 for the remaining testing pairs.

\subsection{Performance of the water model}
\begin{figure}[H]
\includegraphics[width=\linewidth]{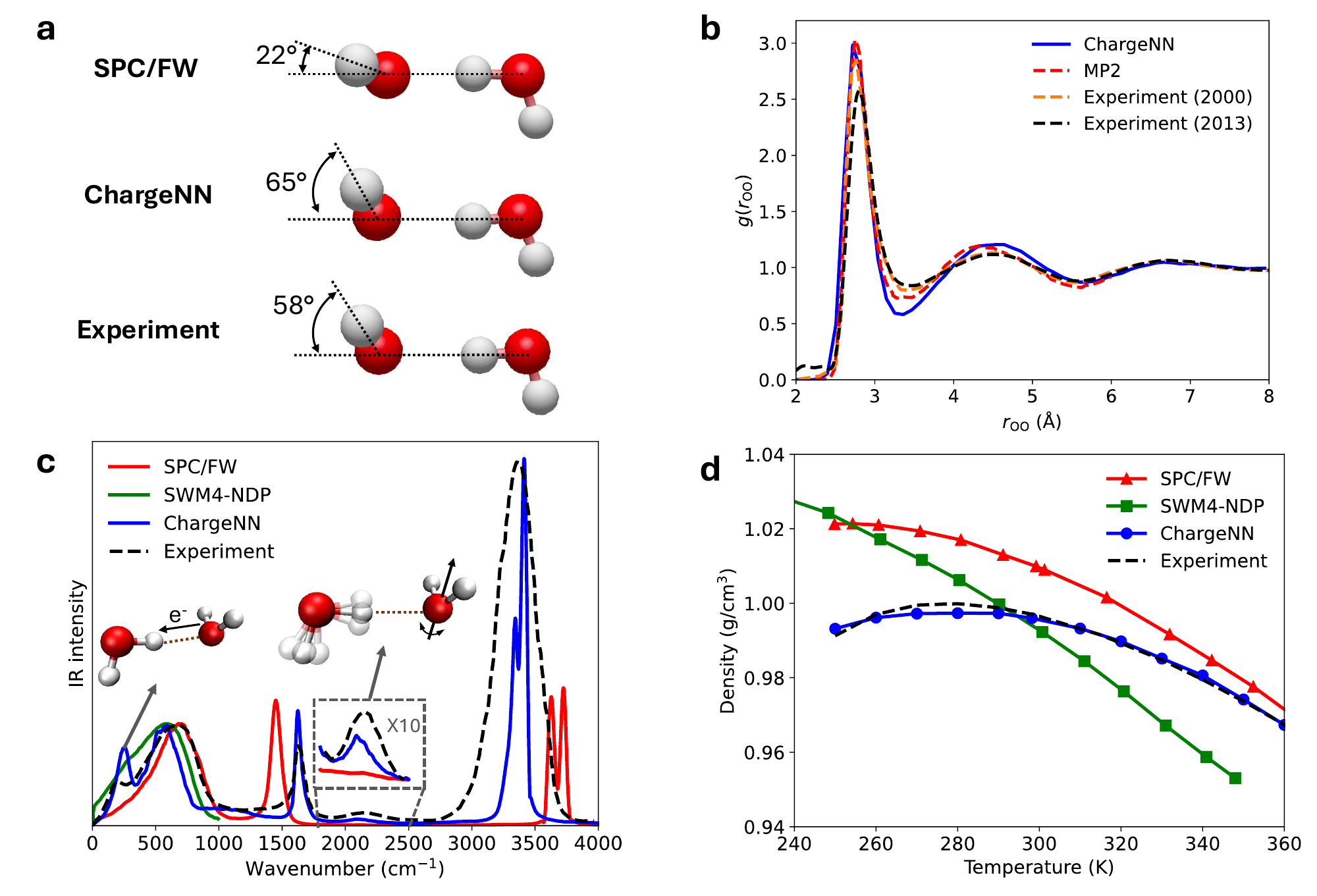}
\centering
\caption{\textbf{Assessment of the ChargeNN water model.} (\textbf{a}) Geometries of water dimers optimized by SPC/FW and ChargeNN, compared to the experimental structure\cite{dyke1977structure}. (\textbf{b}) The liquid radial distribution functions of the oxygen-oxygen distance obtained with ChargeNN, MP2\cite{del2013bulk} and experiments~\cite{sorenson2000can,skinner2013benchmark} under ambient conditions. (\textbf{c}) Liquid infrared spectra generated by SPC/FW, SWM4-NDP\cite{han2023incorporating}, ChargeNN and experiment\cite{max2009isotope} under ambient conditions. (\textbf{d}) Temperature dependent densities from SPC/FW\cite{valle2024accuracy}, SWM4-NDP\cite{yu2013six}, ChargeNN and experiment\cite{kell1975density} at 1 atm.}
\label{fig:water_model}
\end{figure}

\begin{table}[ht]
\centering
\caption{\textbf{Benchmark results for the ChargeNN water model}, compared to SPC/FW, SWM4-NDP, and experiments.}\label{tab:bench_water}
\begin{tabular}{llllll}
\hline\hline
	&	Units	&	ChargeNN	&	SPC/FW$^c$	&	SWM4-NDP$^d$	&	Expt.$^e$	\\
\hline											
\textbf{Monomer}											\\
$\mu_\mathrm{total}$	&	Debye	&	1.79	&	2.19	&	1.85	&	1.85	\\
\textbf{Dimer}	&		&		&		&		&		\\
$\mu_\mathrm{total}$	&	Debye	&	2.397	&	3.594	&	2.062	&	2.643	\\
$r_{\mathrm{OO}}$	&	Angstrom	&	2.81	&	2.73	&	2.83	&	2.98	\\
$\theta_{A}$	&	degree	&	65	&	22	&	71	&	58	\\
$E_{\mathrm{int}}$	&	kcal/mol	&	-4.92	&	-7.14	&	-5.15	&	-5.4	\\
\textbf{Liquid}											\\
$\left< r_{\mathrm{OO}} \right>$	&	Angstrom	&	0.97	&	1.03	&	0.96	&	0.97	\\
$\left< \angle_{\mathrm{HOH}} \right>$	&	degree	&	106.1	&	107.7	&	104.52	&	106.5	\\
$\rho$	&	g/cm$^3$	&	0.996	&	1.012	&	0.998	&	0.997	\\
$E_{\mathrm{int}}$	&	kcal/mol	&	-9.917	&	-11.926	&	-9.927	&	-9.92	\\
$H_{\mathrm{vap}}$	&	kcal/mol	&	10.50	&	10.72	&	10.52	&	10.52	\\
$D$	&	10$^{-5}$ cm$^2$s$^{-1}$	&	2.08	&	2.32	&	2.33	&	2.3	\\
$\left< \mu_{\mathrm{mol}} \right>$	&	Debye	&	1.76$^a$	&	2.39	&	2.46	&	2.9	\\
$\epsilon$	&		&	50.9$^b$	&	79.6	&	79.0	&	78.4	\\
\textbf{Ice}											\\
$T_{\mathrm{melt}}$	&	K	&	288	&	190	&	185	&	273	\\
\hline\hline
\multicolumn{6}{p{14cm}}{\footnotesize
$^a$The origin was set to be the center of the nuclear charge of each molecule. $^b$Molecules were wrapped to the unit cell.
$^c$The gas-phase properties, interaction energy and melting temperature for SPC/FW were computed in this work, while other data were from ref.\citen{wu2006flexible}. 
$^d$SWM4-NDP results from ref.\citen{lamoureux2006polarizable,wang2022flexible}. 
$^e$Experimental results from ref.\citen{dyke1973electric,dyke1977structure,jancso1974condensed,krynicki1978pressure,fernandez1995database,badyal2000electron}.
}
\end{tabular}
\end{table}
To assess the performance of the ChargeNN water model, we calculated various properties of water in gas, liquid, and solid states. Computational details are provided in the SI. We compare our results with those of SPC/FW\cite{wu2006flexible}, which is the fixed charge counterpart of the 3-site ChargeNN model, and those with a widely used polarizable model SWM4-NDP\cite{lamoureux2006polarizable}.  As shown in Table \ref{tab:bench_water}, the ChargeNN's predictions of water monomer and dimer properties agree well with experimental results, demonstrating its capability to accurately simulate gas-phase water clusters. Figure \ref{fig:water_model}\textbf{a} illustrates that the angle $\theta_{A}$ of the water dimer obtained with ChargeNN (65$^{\circ}$) closely matches the experimental value of 58$^{\circ}$, in contrast to the unreasonable 22$^{\circ}$ by SPC/FW, highlighting the significant impact of polarization and charge transfer on the equilibrium geometries. Despite of the inclusion of intermolecular charge-transfer interactions in our model, enhanced electrostatics using high-order polarization may further improve the structure predictions.\cite{cardamone2014multipolar} For instance, the O-O distance of the water dimer optimized by the multipolar AMOEBA (2.91 \AA)\cite{laury2015revised} aligns more closely with experimental measurement (2.98 \AA) compared to ChargeNN (2.81 \AA).

Table \ref{tab:bench_water} also demonstrates that the ChargeNN water model accurately reproduces bulk water properties at room temperature and standard pressure, such as average bond length and angle, density, interaction energy, vaporization heat, and diffusion constants. As shown in Figure \ref{fig:water_model}\textbf{b}, ChargeNN produces an RDF for liquid oxygen-oxygen distances in good agreement with the MP2\cite{del2013bulk} and early experimental data\cite{sorenson2000can}. Nevertheless, the latest x-ray diffraction experiment\cite{skinner2013benchmark} has measured a less pronounced first peak of approximately 2.57 than the predicted one of 2.98, indicating the need for further parameter refinement to mitigate the ``over-structured'' hydrogen-bond network. The predicted infrared spectra in Figure \ref{fig:water_model}\textbf{c} were obtained through a Fourier transformation of the autocorrelation function of the total dipole moments, in which the one produced by ChargeNN matches the experimental spectrum closely in both band positions and intensities. Notably, ChargeNN successfully captures the hydrogen-bond stretching peak at $\sim$200 cm$^{-1}$, which is absent in the spectra by SPC/FW and SWM4-NDP. Our results and a previous study\cite{han2023incorporating} indicate that accurately describing hydrogen-bond stretching requires precise polarization and intermolecular CT. Additionally, ChargeNN retrieves the bending-libration combination band at $\sim$2150 cm$^{-1}$, which is missing in the SPC/FW spectrum, further underscoring the importance of explicit polarization in predicting vibrational properties. Nevertheless, we observe that the O-H stretching band at around 3400 cm$^{-1}$ is considerably narrower than the reference, which may be improved by replacing the harmonic intramolecular forces with anharmonic forces\cite{shi2013polarizable} or reactive force fields\cite{van2001reaxff, leven2020reactive}. 

We computed the molecular dipole moments by setting the origin as the nuclear charge center of each partially charged water molecule, following the reported protocol\cite{han2021determining, han2023incorporating}. While ChargeNN accurately reproduces dipole moments for water monomer and dimer, as well as the liquid IR spectrum, it yields an average molecular dipole moment of approximately 1.76 Debye, which is notably lower than the experimental estimation of 2.9 Debye. While the choice of the origins may have a significant impact on the dipole moments of charged molecules, there could be parallel problems in the population analysis methods. A recent comprehensive study discovered that none of the selected charge models (Mulliken, Hirshfeld, QTAIM, RESP, ChelpG, Hirshfeld-I, and NPA) is capable of reproducing both total dipole moments and average molecular dipole moments of water clusters.\cite{han2021determining} Similarly, CM5 charges agree well with the reference values for total dipole moments, but fall in short to provide a reasonable average molecular dipole moment. Additionally, our water model underestimates the dielectric constant obtained from the total liquid dipole moments, which is largely due to the non-uniqueness of the dipole moment of an extended system according to the modern theory of polarization\cite{resta2007theory}. 

The accurate characterization across a range of temperatures is important to the performance of a water model. Figure \ref{fig:water_model}\textbf{d} demonstrates that ChargeNN successfully predicts the temperature-dependent density of liquid water, with the temperature of maximum density around 280 K, very close to the experimental value of 277 K. In contrast, this prediction remains a challenge for the fixed charge model SPC/FW and other well-tuned 3-site models\cite{izadi2016accuracy}, emphasizing the substantial improvement resulting from ChargeNN polarization. The polarizable SWM4-NDP also falls short in predicting the density variation with temperature. Additionally, Figure S8 presents that the computed temperature-dependence of vaporization enthalpy by ChargeNN agrees well with the reference. For the simulation of ice, we calculated the orientational tetrahedral order parameters of the equilibrium configurations across various temperatures, starting from a perfect hexagonal ($I_{h}$) ice structure. The ChargeNN yields a melting temperature of 288 K, close to the experimental value of 273 K. While this result is significantly better than SPC/FW at 190 K and SWM4-NDP at 185 K\cite{wang2022flexible}, there is still room to fine-tune the parameters to achieve ice properties comparable to TIP4P/Ice\cite{abascal2005potential}. 

\begin{figure}[H]
  \includegraphics[width=\linewidth]{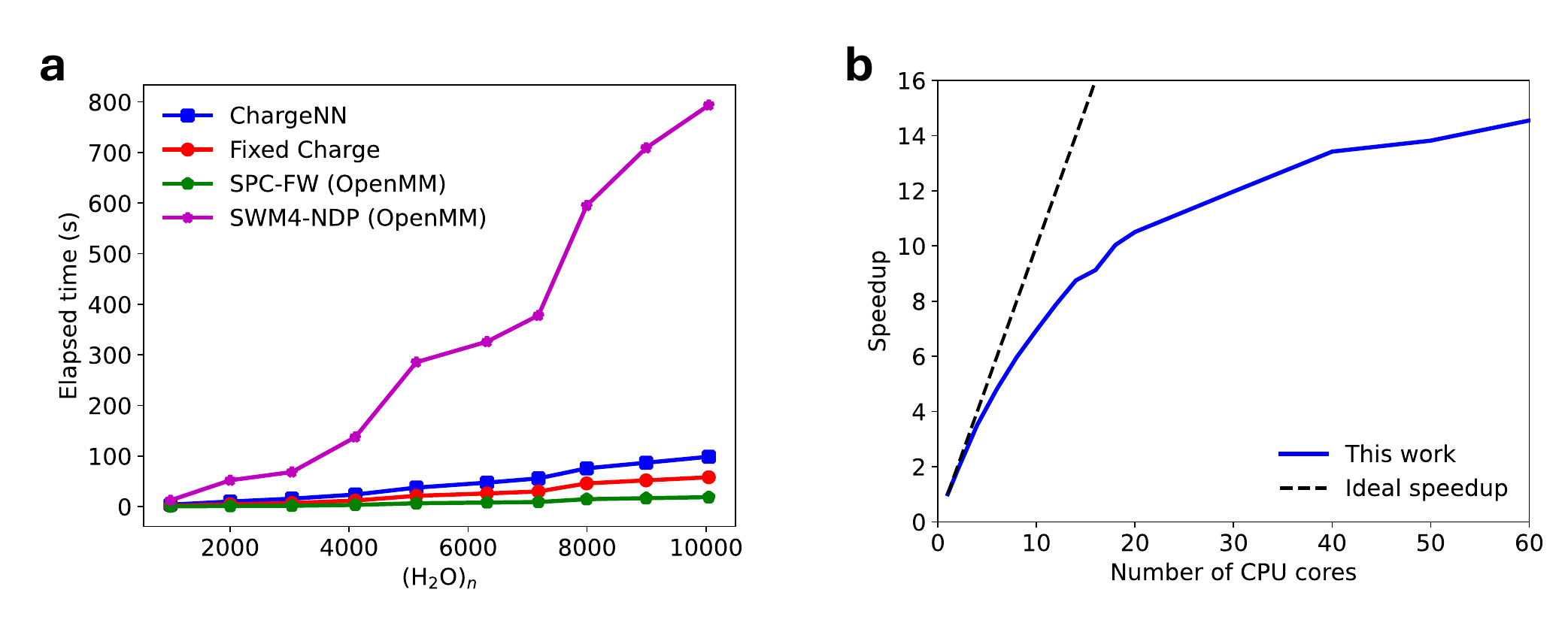}
  \centering
  \caption{\textbf{Timing performance of ChargeNN.} (\textbf{a}) Elapsed time for force calculations on varying water box sizes using a single CPU core (AMD EPYC 7H12, 2.6GHz), comparing in-house ChargeNN and fixed point charges models with OpenMM's SPC-FW and SWM4-NDP. (\textbf{b}) Speedup of ChargeNN calculations of periodic (H$_{2}$O)$_{10044}$ with respect to the number of the CPU cores.}
  \label{fig:time}
  \end{figure}

As demonstrated in Figure \ref{fig:time}\textbf{a}, ChargeNN exhibits exceptional computational efficiency, making it highly promising for large-scale water simulations. For instance, using a single CPU core for a PBC box containing 10044 water molecules, ChargeNN's force calculation time is only one-eighth of that for the economical polarizable SWM4-NDP and merely five times that of the 3-site fixed charge model SPC-FW, both implemented in OpenMM for CPU platforms. Moreover, while the incorporation of machine-learned charges involves additional steps such as feature preparation, charge prediction, and charge derivatives, these ChargeNN-related steps only account for approximately 40\% of the total time. Furthermore, the non-ChargeNN components (equivalent to fixed charge computations) are three times slower than OpenMM's SPC-FW, indicating room for algorithmic improvement. In addition, Figure \ref{fig:time}\textbf{b} reveals suboptimal parallel efficiency of the ChargeNN program, necessitating further optimization. We intend to transfer the algorithm to C++ for enhanced performance, with plans to adapt it to CUDA or OpenCL for extensive parallelization on graphics processing units (GPUs) in future implementations.

\subsection{Demonstrative application}
\begin{figure}[H]
\includegraphics[width=\linewidth]{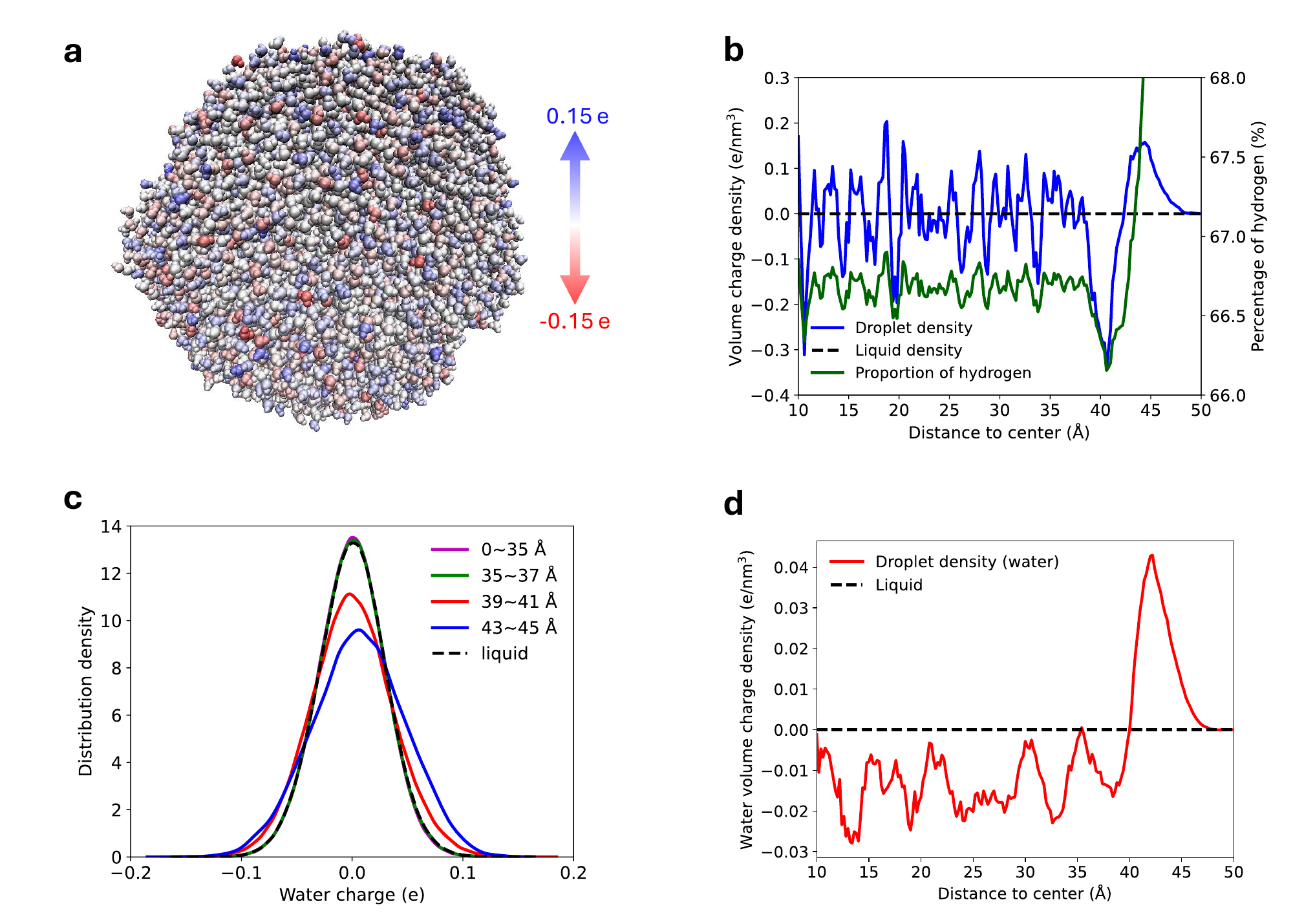}
\centering
\caption{\textbf{Statistical results for the illustrative application with a large droplet.} (\textbf{a}) Geometry of the studied droplet colored according to molecular charges.  (\textbf{b}) Correlation between VCD and PHA as a function of the layer distance to the droplet center.  (\textbf{c}) Charge distributions of water molecules (\textbf{d}) Volume charge density using net molecular charges. The center of mass of each water molecule was used to compute the layer distance to the center.}
\label{fig:app}
\end{figure}

As an illustrative application, we probe the origin of the emerging on-water reactivity on the air/water interface using ChargeNN. The electronic process for creating the strong electric fields on the water microdroplet surface remains elusive, and has been attributed to the abundance of dangling OH bonds at the interface \cite{inoue2020reorientation,pezzotti20172d}, the presence of OH$^-$/H$^+$ due to proton transfer \cite{hao2022can,heindel2022spontaneous, heindel2024role} and electron transfer \cite{poli2020charge,ben2022electric,liang2023water}. To understand the source of the interfacial electric fields, we carried out statistical analysis to identify differences in geometries and charge distributions between the droplet and bulk liquid. We extracted the last 100 structural snapshots at 100 femtosecond intervals from a 1-nanosecond (ns) MD trajectory of a large water droplet with a radius $\sim$45 {\AA} containing 10327 water molecules. For the liquid simulation, we analyzed the last 500 snapshots from a 1-ns MD trajectory of a large unit cell containing 2123 water molecules. 

Our results reveal that the primary factor influencing the layer charge density is the hydrogen-to-oxygen ratio. The volume charge density (VCD) has been extensively used as an indicator for the strength of the layer electric fields\cite{poli2020charge}, and is highly correlated with the proportion of the hydrogen atoms (PHA) (see Figure \ref{fig:app}\textbf{b}). For deep layers inside the droplet at distances less than 37 {\AA} from the center, the PHA fluctuates around an equilibrium percentage of 67\% indicating an homogeneous distribution, with the VCD oscillating near the liquid VCD of 0 e/nm$^3$. The PHA on the intermediate layer at the distance of 41 {\AA} from the center drops significantly to 66.2\% , and coincides with the most negatively charged layer with a VCD of -0.35 e/nm$^3$. Beyond this layer, the PHA grows continuously to break the theoretical limit of 66.7\% (that is, 2 H atoms and 1 O atom per water) and even exceed 68\% at the air/water interface, revealing an inhomogeneous distribution of H-bond network near the surface, where the VCD increases consistently and reaches a maximum of 0.18 e/nm$^3$ at 45 {\AA}. Our results align with the ``dangling OH theory'' \cite{inoue2020reorientation,pezzotti20172d} that more OH bonds dangle on the droplet surface due to the partial collapse of the hydrogen-bond network near the surface, leading to a higher proportion of hydrogen atoms and a positively charged surface layer. Consequently, the inward layer near the interface is abundant in oxygen atoms, and overall negatively charged. 

In addition, the considerable charge transfer also significantly contributes to the layer charge density. As shown in Figure \ref{fig:app}\textbf{c}, in the inner shells at distances shorter than 37 {\AA} from the center, the molecular charges are closely clustered around zero, similar to the distribution in the liquid phase. In contrast, the vast interfacial charge separation results in a substantially increasing population of the charged water molecules.  For understanding the explicit CT effect on the charge density, we obtained the water VCDs using net molecular charges of the water molecules whose centers of mass reside within the layers, as demonstrated in Figure \ref{fig:app}\textbf{d}. The interior of the droplet is overall negatively charged with a water VCD of $\sim$-0.015 e/nm$^3$, indicating electron migration from the droplet surface to the inter layers. This is drastically different from the uniform charge density distribution in liquid phase. Furthermore, the water VCD of the positive droplet surface is approximately 0.045 e/nm$^3$, constituting 25\% of the total VCD at the air/water interface.
\section{Application Outlook}
Beyond the simulation of water, we can envisage broader applications with ChargeNN. Firstly, the ChargeNN water model can be seamlessly combined with the existing force fields to simulate interactions between proteins and water solvents. Such interactions markedly affect the structures and dynamic behavior of proteins\cite{prabhu2006protein}. Additionally, integrating chargeNN water model with QM electron structures via a QM/MM interface will allow accurate descriptions of outer solvent shells for reactive centers. Moreover, incorporating empirical valence bond (EVB)\cite{warshel1980empirical} and reactive force fields (ReaxFF)\cite{van2001reaxff, leven2020reactive} into ChargeNN will enable simulations of bond breaking and formation, facilitating efficient modeling of proton transfers through hydrogen-bond networks and chemical reactions. Furthermore, accurate simulation of protein-ligand electrostatic interactions requires incorporating electronic polarizability, which induces screening effects that weaken electrostatics in the buried environment.\cite{rocklin2013blind} A recent study also shows that QM electronic polarization is essential for accurately producing spectral densities for proteins.\cite{zuehlsdorff2019influence} Therefore, we aim to extend the ChargeNN to other atoms and molecular systems, enabling generalizable predictions of atomic charges across diverse local environments and achieving QM-level charge assignments for complex macromolecules. Combining accurate charges and force fields will allow for the depiction of instantaneous polarization and transient charge transfer in molecular dynamics. Coupled with well-designed energy functionals, fast prediction of \textit{ab initio} polarization from neural networks would significantly enhance the accuracy of atomistic simulations for large chemical systems other than water.

\section{Conclusions}
Modeling water has been pivotal task in theoretical chemistry, aiming at understanding its unique thermodynamic and electronic characteristics and propensities. Nevertheless, simulating polarization remains a challenge due to the volatile nature of electron distribution and its high sensitivity to local electric fields, which necessitates quantum mechanical treatments for accurate descriptions. Despite significant advancements in low-scaling QM electronic structures and advanced computing technology, QM calculations for large water systems remain prohibitively expensive. Machine learning offers promise for overcoming the cost obstacles and predicting atomic charges with QM accuracy. However, the existing ML charge models are limited to static calculations of specific systems, incapable of simulating polarization in both liquid and gas phase water during the structural and dynamical evolution. 

In this work, we address the aforementioned problems by introducing a dynamic and polarizable water model that leverages machined-learned MP2-level partial charges. The charge model, termed ChargeNN, employs deep neural networks to map the interaction classified functions that accurately characterize the local environment onto the CM5 charges computed with MP2 electron density, demonstrating high accuracy and generalizability. Equipped with quantum atomic charges and charge transfer predicted by the neural networks, the ChargeNN water model successfully reproduces a variety of water properties across different temperatures and phases in excellent agreement with the experimental measurements, validating its capacity to generate electronic and thermodynamic quantities via fast and long molecular dynamics.

Additionally, we probed the origin of the strong local electric fields on the droplet surface by conducting molecular dynamics of both liquid water and a large droplet using the ChargeNN water model. The findings clearly indicate that the layer electric fields on the droplet surface are primarily induced by breakage of the hydrogen-bond network, which leads to distinct proportion of hydrogen atoms from the liquid phase. Furthermore, surface-to-interior charge transfer also considerably contributes to the layer electron density. 

\appendix

\section*{Appendix}
\subsection*{MBE(3)-OSV-MP2 electronic structure}
 The atomic charges were calculated using the MBE(3)-OSV-MP2 method. The OSV approximation has demonstrated a significant reduction in the computational effort required for correlated methods without considerable loss in accuracy.\cite{yang2011tensor,kurashige2012optimization,yang2012orbital,zhou2019complete,liang2021third,ng2023low,yang2024making} The OSVs $(\mathbf{Q}_{k})$ can be obtained through a simple singular value decomposition of the diagonal pair MP2 amplitudes $\mathbf{T}_{kk}$:
 \begin{equation}
[\mathbf{Q}_{k}^{\dagger}\mathbf{T}_{kk}\mathbf{Q}_{k}]_{\overline{\mu}_{k}\overline{\nu}_{k}} = \omega_{\overline{\mu}_{k}}\delta_{\overline{\mu}\overline{\nu}}.
 \end{equation}
The inherent sparsity allows the OSV space to be reduced by setting a cutoff for the singular values $\omega_{\overline{\mu}_{k}}$, which indicates the significance of OSVs. To further address the computational scaling challenge, a many-body expansion of OSV-MP2 amplitudes and density matrices with orbital-specific partitioning was introduced:
\begin{equation}
    \mathbf{T}_{(ii, ii)} = \mathbf{T}^i_{(ii,ii)}  + \sum_{k}\Delta\mathbf{T}^{i,k}_{(ii,ii)} 
                        + \sum_{k>l}\Delta\mathbf{T}^{i,k,l}_{(ii,ii)}.
\end{equation}
The population analysis can be performed using the total relaxed one particle density matrix, which can be computed as the sum of the density matrices of Hartree Fock, relaxed MP2 and molecular orbital (MO) relaxation:
\begin{equation}
    \mathbf{P} = \mathbf{P}^{\mathrm{HF}} + \mathbf{P}^{\mathrm{MP2}} + \mathbf{P}^{\mathrm{MO}}.
\end{equation}
The relaxed OSV-MP2 density matrix in MO basis can be obtained with
\begin{equation}
    \mathbf{P}^{\mathrm{MP2}} = \sum_{ij}\mathbf{T}_{ii}\left< \mathbf{X}^{\mathrm{T}}_{ij} + \mathbf{X}^{\bot}_{ji} \right>,
\end{equation}
where $\mathbf{T}_{ii}$ denotes the diagonal semi-canonical amplitudes, while $\mathbf{X}^{\mathrm{T}}_{ij}$ and $\mathbf{X}^{\bot}_{ji}$ represent the upper and lower diagonal blocks of the matrix containing OSV-MP2 responses.\cite{zhou2019complete}

Additionally, we developed a large-scale parallelism strategy utilizing efficient memory management and data communication, which enables computations of MP2-level charges for large molecular systems.\cite{liang2021third}

\begin{acknowledgement}
The authors acknowledge financial supports from the Hong Kong Research Grant Council (17309020, 17305724), Hong Kong Quantum AI Lab through AIR@InnoHK program of Hong Kong Government, and the Hung Hing Ying Physical Sciences Research Fund of the University of Hong Kong. Q.L. acknowledges Dr. Xinyan Wang and Dr. Ruiyi Zhou for valuable discussions.
\end{acknowledgement}

\begin{suppinfo}
The Supporting Information is available free of charge. The SI document comprises the results of the convergence of atomic charges and charge transfer with respect to atomic distances, testing MAEs relative to the training set size, parity graphs of molecular charges and charge transfer, optimized parameters for the ChargeNN water model, computational details for benchmarking ChargeNN, and some supplementary benchmark results.

\end{suppinfo}

\bibliography{manuscript}

\end{document}